\documentclass[aps,pre,subeqns.floatfix,twocolumn,amsmath,showpacs]{revtex4-1}
\usepackage[pdftex]{graphicx}

\usepackage{amsmath}
\usepackage{verbatim}
\usepackage{mathtools}
\usepackage{amsfonts}
\usepackage{epstopdf}
\usepackage{graphicx}
\usepackage{subfigure}
\usepackage{hyperref}
\usepackage{color}
\usepackage{dcolumn}
\usepackage{bm}
\usepackage{mathrsfs}
\usepackage[normalem]{ulem}
\DeclarePairedDelimiter{\ceil}{\lceil}{\rceil}
\begin{document}
\title{Effect of repulsive links on frustration in attractively coupled networks} 
\author{Sayantan Nag Chowdhury}
 \author{Dibakar Ghosh}
 \author{Chittaranjan Hens}
\affiliation{Physics and Applied Mathematics Unit, Indian Statistical Institute, 203 B. T. Road, Kolkata-700108, India}
\date{\today}
\begin{abstract}
 We investigate the impact of attractive-repulsive interaction in networks of limit cycle oscillators. Mainly we focus on the design principle for generating an anti-phase state  between adjacent  nodes in a complex network.   
 We establish that a partial  negative control throughout the branches of  a spanning tree inside the positively coupled limit cycle oscillators works efficiently well in comparison with randomly chosen negative links  to establish zero frustration (anti-phase synchronization) in bipartite graphs. Based on the emergence of zero frustration,
we  develop a universal  
  $0-\pi$
  rule to understand the anti-phase synchronization in a bipartite graph. Further, this rule is used  to construct a non-bipartite graph  for a given non-zero frustrated value.  We finally show the   generality of $0-\pi$ rule by  implementing it in  arbitrary undirected non-bipartite graphs of attractive-repulsively coupled limit cycle oscillators and successfully calculate the non-zero frustration value which matches with numerical data. The validation of the rule is  checked through the bifurcation analysis of small networks.
  Our work may unveil the underlying mechanism of several synchronization phenomena that exist in a network of oscillators having a mixed type of coupling. 
\end{abstract}

\pacs{}
\maketitle

\section{Introduction}

 A tug-of-war between attractive and repulsive coupling can establish complex behavior \cite{Sayntan_2019_NJP,  Hong_2011_PRL,Zanette_2005_EPL, Tessone_2008_EPJB,Solitary_in mixed environment,Martins_2011,Volcano_transition,
 	Soumen_2019_Chaos,
sir1} in a large class of coupled oscillators. For instance,  it can create resonance in a network of spin like dynamical variables \cite{Tessone_2009_EPJB}
 ~ or helps in  entertainment process  for biological  networks \cite{Biological systems,leyva-irene2006,perc-plosone 2011}. A careful selection of negative links of attractively coupled oscillatory system can systematically establish homogeneous or inhomogeneous solution \cite{Bera2016,Hens2014,Hens_2013,Gade2013,Kundu_2019_Chaos,Dixit2019} or may generate diverse chimera states ranging from clustered death to amplitude chimera states \cite{Sathiyadevi_2019,Sathiyadevi_2018, Sathiyadevi_2018(2), Mishra2015}.  On the other hand,  a   signed (mixture of attractive-repulsive coupling) and balanced graph of phase oscillators    may uncover the underlying structure of the networks \cite{Wu2016} on the basis of synchronous phase clusters.  Identification of  critical nodes as well as graph partitioning  in  real ecological signed graphs can also be recognized from the reorganization of synchronized clusters \cite{Bacelar2014,Giron2016}. 
   Also, the network of Kuramoto dynamics \cite{Kuramoto2005} having phase lag \cite{Sakaguchi1986} in identical frequency environment can split the stationary phases into several clusters \cite{Nicosia2013} revealing the underlying symmetry of the given network. On the other hand, the inhibition or delayed feedback naturally suppresses (controls)   the complex synchronization pattern \cite{Khanra2018,Timms2014,Louzada2012,Louzada2013, Lohe2015} or generates multistability \cite{Labavic2017}. Besides,  rewiring links in a network  may reveal anti-phase states between the adjacent nodes in repulsively coupled phase oscillators \cite{Levnajic2011,Levnajic2012}. 
 \par   However, the emergence of anti-phase state \cite{pikovsky} between adjacent nodes in attractive-repulsively coupled complex networks is not well explored, till now to the best of our knowledge. Here, we would like to understand the suitable characteristics of a network, for which the adjacent nodes of that network can settle themselves into anti-phase state  in the presence of tiny fraction of repulsive links of a group of attractively coupled limit cycle oscillators and in absence of such attribute, the network cannot reveal anti-phase states.  
 
 \par  The manifestation of anti-phase synchronization has been profoundly studied in various physical systems, both from numerical and experimental \cite{Fox_2007, Lewis_2009} point of view. One of the premier specimens, to observe such an emergent phenomenon, is the cortical neural network \cite{Belykh_2018,Li2011}. Anti-phase patterns have been found there too. Recently, anti-phase collective synchronization \cite{Sebek2019} has been observed among two pairs of electrochemical oscillators with strong internal and weak external coupling. All these inspections attest to the necessity of studying anti-phase synchronization in coupled complex networks. In this article, based on the emergence of anti-phase synchronization, we would like to classify the entire class of networks into two distinct groups of classes, one which  adjacent nodes can exhibit such anti-phase patterns, while the   oscillators of the other types of networks fail to rearrange into such anti-phase formation.

 The essential term to quantify the anti-phase synchronization between two neighboring nodes $i$ and $j$ is to calculate the local frustration between them and which is express as  
   $ f_{local}^{ij}= C_{ij}\left(1+\cos(\phi^i-\phi^j)\right)$,
where 
 $C_{ij}$ 
is the elements of the symmetric binary connectivity matrix $C$,
$i$ 
is an arbitrary node in the network, 
$j$ 
is the neighbor of 
$i$ and $\phi^i$
 is the intrinsic phase 
 of the 
 $i$th node. Here,
 $ 0 \leq f_{local}^{ij} \leq 2$.   
  $L$
  is the total number of  links, then  
  $F=\Big\langle \frac{1}{L}\sum_{i<j}^{}f_{local}^{ij} \Big\rangle_t$ 
 represents the total frustration in the entire network and it accumulates all 
 $f^{ij} _{local}$ 
 present in the network. Here ${\langle \cdot \cdot \cdot \rangle}_t$ stands for time average.  
  $F=0$, i.e., non-frustrated system
 implies the anti-phase states in the entire networks and thus, each pair of adjacent nodes will follow anti-phase 
 ($|\phi^{i} - \phi^{j}| =\pi$) 
 states between them and the dynamics of the system will no longer be able to stretch along links beyond the phase differences of $\pi$. 
 $F=2$ signifies in-phase synchrony in the network as $|\phi^{i} - \phi^{j}| =0$. The non-zero value  of 
 $F$  reflects frustration nature \cite{Levnajic2011,Levnajic2012,Levnajic2019,Sathiyadevi_2019,
 Frustration,Zanette_2005_EPL,Sen_2017_PRE} of the system, in which most of the adjacent nodes will be out-of phase to each other, as at least few links will be squeezed to a phase difference
less than $\pi$ and frustrate the simultaneous minimization
of the energy of all the interacting pairs of subunits of a system. Treating $F$ as the elastic potential energy \cite{Zanette_2005_EPL} contained within the connected network, $F$ captures how squeezed a link is.
 We examine here, how  attractive-repulsively coupled oscillators  can impact
 $F$ 
 in a complex  network. We show here that   a few number of negative links may induce zero frustration in a bipartite graph.   Therefore, the main focus of this paper is to answer the following key questions:  (i) which links are  crucial to  obtain anti-phase states  ($F=0$) in a bipartite graph?  More precisely, we would like to identify which positive links should be replaced with negative ones for faster decay of $F$ from $2$ to $0$?  And secondly, (ii) can we design a non-bipartite graph which can give us as our expected desired $F$  ($F\neq 0$)? \\
 
 \par Rest of the paper is arranged as follows: in Sec. II, we discuss the model description and how the attractive and repulsive links are chosen in the considered network. In Sec. III, we show the efficient arrangement of negative links in bipartite graph of attractively coupled limit cycle oscillators. The arrangement can produce zero frustration easily.
Sections IV and V analyze the bifurcation of the oscillatory nodes and establish a $0-\pi$ rule to understand the zero frustration in bipartite graph.
In Sec. VI, non-bipartite graph ( based on $0-\pi$  rule) are constructed for desired nonzero $F$. Finally, we show that our proposed rule can be  implemented for any non-bipartite graph and we can easily calculate $F$ for any arbitrarily chosen graph of limit cycle in presence of attractive-repulsive coupling in Sec. VII. Lastly, we conclude our findings in Sec. VIII.

\section{Mathematical model}

We start with $d$-dimensional 
  oscillatory agent. The dynamics of each agent is used  over the top of  each node  of the network. The evaluation of $i$-th agent  is captured by

     \begin{equation} \label{eq:1}
 	\begin{array}{lcl}
 	
 	{\bf \dot{x}}^i=\psi({\bf x}^i)+K_A \sum_{j=1}^{N} A_{ij} H({\bf x}^i,{\bf x}^j)\\
 	+K_R \sum_{j=1}^{N} B_{ij} H({\bf x}^i, {\bf x}^j),i=1,2,...,N
 	\end{array}
 	\end{equation}	
 with the vector ${\bf x}^i (t)$ is the state variable$, ~\psi: \mathbb{R}^d \rightarrow \mathbb{R}^d $ 
 given by the system's intrinsic dynamics,  
  the attracting coupling strength
  $K_A > 0$ 
 ensures in-phase
  synchronization 
 for a certain threshold and 
 the repulsive (negative) strength
 $K_R$ 
 tries to push the entire system  out of phase synchronization by disturbing  the in-phase synchronization manifold. The function
 $H({\bf x}^i, {\bf x}^j)$ 
 is the vector coupling representing diffusive type,  
 i.e., $H({\bf x}^i,{\bf x}^j)=({\bf x}^j-{\bf x}^i)$. Here
 $A_{ij}$ and $B_{ij}$ 
 are the binary elements encoding the information of the underlying attractive and repulsive connectivity patterns among the dynamical agents. The total number of links of the entire network   (matrix $C$) is the sum of links 
 encoded in the matrices $A$ and $B$. 
 As a paradigmatic model, we  choose Landau-Stuart (LS) oscillators  where the state dynamics of each limit cycle oscillators is represented by 
 
 \begin{equation} \label{eqL}
 \psi({\bf x}^i)=\left(
 \begin{array}{c}
 \left[1-\left({x^i}^2+{y^i}^2\right)\right]x^i-\omega^i y^i\\\\
 \left[1-\left({x^i}^2+{y^i}^2\right)\right]y^i+\omega^i x^i\\
 \end{array}
 \right), \\
 \end{equation}
  where $H({\bf x}^i,{\bf x}^j)=[{x}^j - {x}^i, 0]^T$ 
  is  the coupling function and 
  $\omega=\omega^i=3$ 
  is identical intrinsic frequency of each node. The intrinsic and instantaneous phase 
  of each node  is calculated by
  $\tan\phi^i= \frac{y^i}{x^i}$.

\par
\begin{figure*}[ht]
	\centerline{
    \includegraphics[scale=0.25]{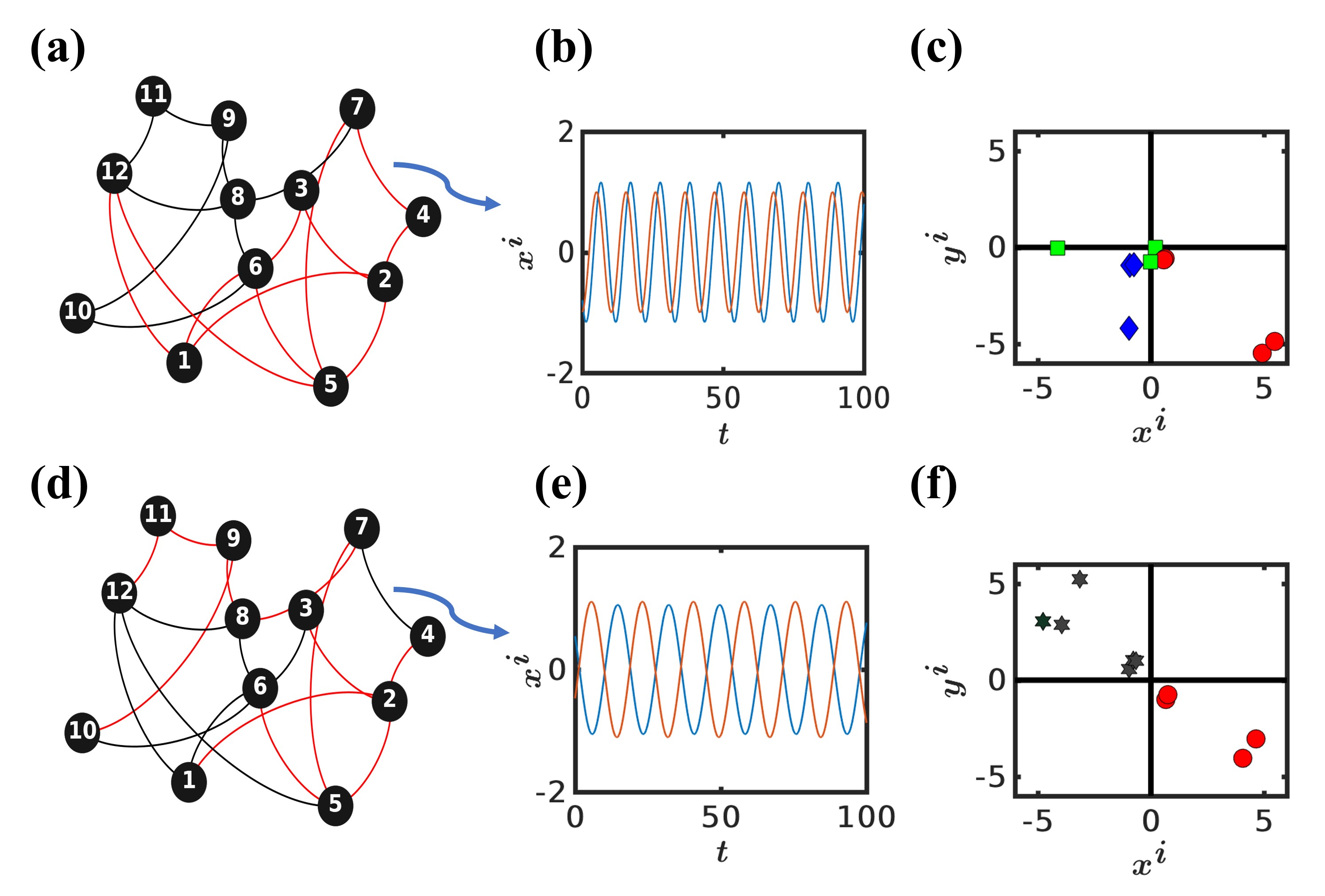} 	}
	\caption{{\bf Impact of different arrangement of repulsive links. } A network with $N=12$ nodes and 
	$L=19$ 	links are considered in (a) and (d) where the black lines are attractive links,  the red (grey) ones are the repulsively coupled ($N_L=11$) 
	links. The underlying   architecture 
	 remains same for both. Due to different choices of repulsive links, same 
	$K_A$ and $K_R$ lead to different scenarios shown in (b), where node $4$ (time series in blue line) and $7$ (time series in red line) are slightly lagged to each other, however they are opposite to each other (anti-phase) shown in  (e). 	For more higher coupling, the entire network reaches to inhomogeneous steady states portrayed in (c) and (f). In the cases of (e) and (f), the frustration  index $F$ is zero but, $F > 0.3$ for  (b) and (c). 
	 $K_A=0.001$ and $K_R=-0.1$ 
	 are chosen for  (b) and (e). 
	 $K_A=0.001$ and $K_R=-4$ 
	 are taken for  (c) and (f). }
	\label{fig1}
\end{figure*}

 To address the first question raised in the introduction,  we start with a small network 
 (Fig.\ \ref{fig1}(a) or \ref{fig1}(d)), having the size 
 $N=12$ 
 nodes 
 and total links
 $L=19$. 
 To test the impact of attractive-repulsive interactions, we introduce negative coupling 
 ($K_R<0$) 
 on randomly chosen 
 $N_L=11$ (approximately $60\%$  
 of the total links) 
 links shown by red color in Fig.\ \ref{fig1}(a). The remaining 
 $L-N_L$  
 links (black arcs) are coupled through weak positive interaction 
 $K_A$.  
 
 If we introduce positive strength over  all links,  the coupled oscillators will oscillate with a common rhythm (not shown here) by adjusting themselves with common phase even with identical amplitude. However, if we  replace 
$N_L (=11)$ 
attractive links by negative strength  
($K_R=-0.1$) (red links), 
the coherent phase distribution is scattered  and out of phase emerges in the system. For instance, the nodes connected for an    arbitrarily chosen  link (link between nodes $4$ and $7$)       reflects the lag synchronization  with slightly distorted amplitude fluctuations between them shown in  Fig.\ \ref{fig1}(b).   Keeping  entire links structure fixed, if we increase the negative strength 
$K_R=-4$, 
the population reaches to steady states (Fig.\ \ref{fig1}(c)) and it 
neither shows in-phase nor anti-phase 
synchronization as all the emerging fixed points (FPs) are either in the third or fourth quadrant of the phase space.
The phase differences between the existing links are not in the difference of $\pi$. Therefore, 
$F$ 
gives a  positive nonzero value 
$0.3$ 
for both cases (Figs.\ \ref{fig1}(b) and \ref{fig1}(c)). 
 This signifies that the links are either in in-phase mode or out-of-phase mode  leading to the existence of at least few frustrated links with $f_{local}^{ij} > 0$ and hence,  weak frustration ($F>0$)   in the entire network.
A distinct and qualitatively different feature ($F=0$) 
appears for another arrangement of negative links (Fig.\ \ref{fig1}(d) in red color), in which any arbitrary node stays in anti-phase  with its nearest neighbor either in a lower negative coupling (Fig.\ \ref{fig1}(e) shown for the node 
$4$ and its neighbor $7$) or in higher negative strength  (Fig.\ \ref{fig1}(f)).  In the higher negative strength, the FPs stay opposite to each other in the phase space (second and fourth quadrants).   The obtained results give us a hint that proper choices of  negative links in the network is essential for constructing a  network with zero frustration, i.e.\, an emergence of anti-phase synchronization ({$f_{local}^{ij}=0$}) between two adjacent nodes, an interesting impact of negative links in positively coupled networks never explored before to the best of our knowledge. The objective of this article  is to link the network structure with the dynamical process such a way that we can understand the underlying mechanism of obtaining 
  zero ($F=0$) and non-zero frustrations ($F>0$)
of positive-negatively coupled environment of limit cycle oscillators with fixed point and oscillation states.

 \section{ Efficient arrangement of repulsive links  for obtaining  anti-phase ($F=0$) in a bipartite graph} 

 To proceed further,  we first focus on our network described in Fig.\ \ref{fig1}. The underlying network is  bipartite by nature. A graph 
$G$ 
is called bipartite if its vertex set 
$\mathcal{V}$ 
can be decomposed into two disjoint subsets $U$ and $V$ 
such that every link in 
$G$ 
joins a node in 
$U$ with a node in 
$V$, i.e., $U~\cap~V=\phi$, where $\phi$ 
is the null set and $U~\cup~V=\mathcal{V}$.
 The nodes of the network described in Fig.\ \ref{fig1} can be splitted into two disjoint sets: 
$U=\{1,3,4,5,8, 10,11\} $ and 
$V=\{2,6,7,9,12\}$. 
The comparative observation in Fig.\ \ref{fig1} 
helps us to conclude that repulsive links, chosen in the Fig.\ \ref{fig1}(d), is somehow more  influential (effective) than other choices of random  repulsive links (Fig.\ \ref{fig1}(a)).  We also observe  that  the repulsive links, chosen in the Fig.\ \ref{fig1}(d), is the minimally connected subgraph which touches  all the nodes  manifested as a spanning tree in the given network. However,
the randomly chosen links in Fig.\ \ref{fig1}(a) cannot form  a spanning tree 
(as the links fail to cover all the nodes) and
$F$ 
never reaches to the minimum value 
$F\sim0$.  
\begin{figure*}[ht]
	\centerline{
    \includegraphics[scale=0.25]{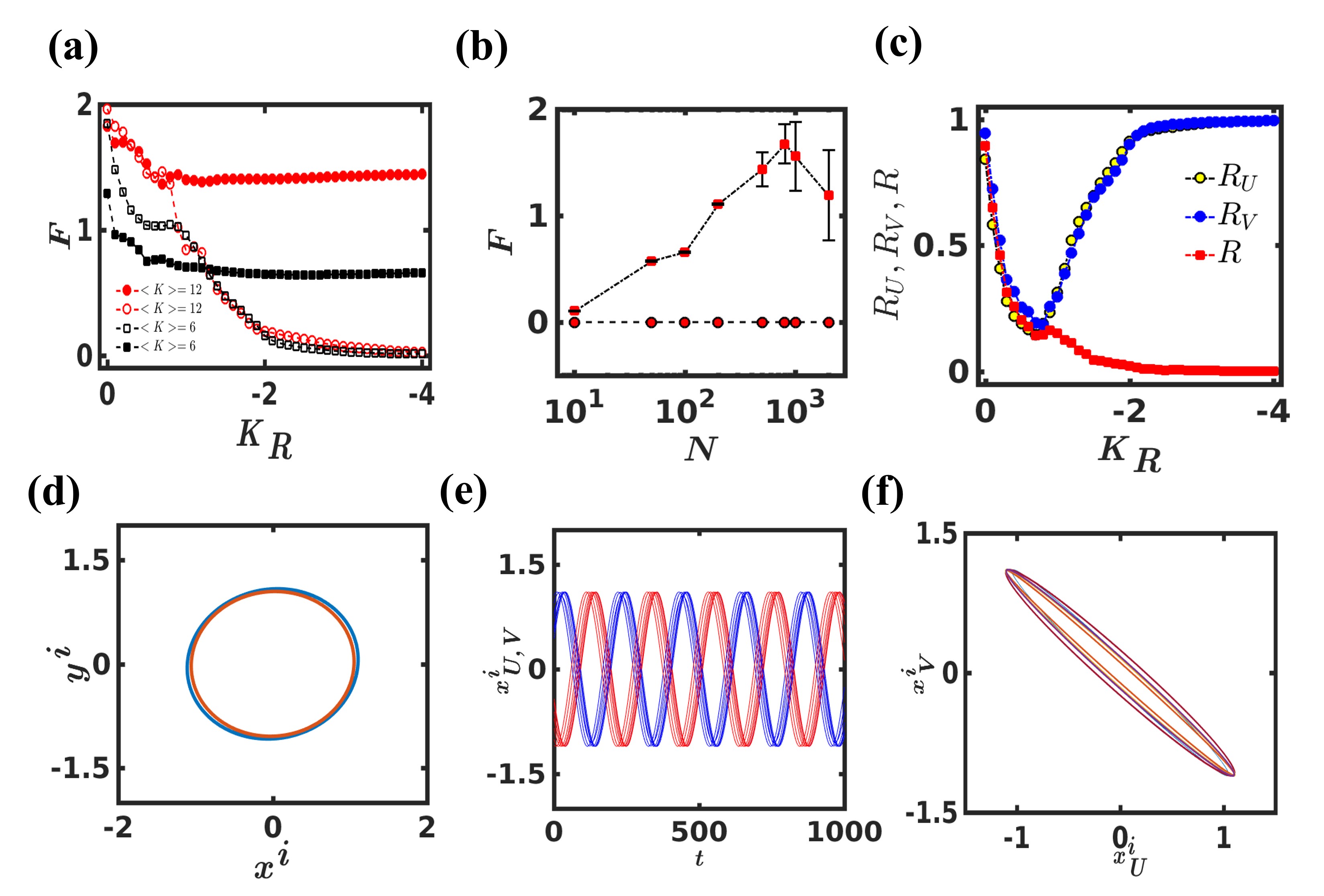}}
	\caption{{\bf Comparison of  frustration index where links are chosen by randomly and through spanning tree. } (a) Two networks (Net1 and Net2) of same size  (N=100) are considered. Red and black  filled circle are drawn by considering randomly chosen ($N-1$) links. Impact of  negative spanning tree is shown with unfilled circles and squares. (b)  Spanning tree creates zero frustration in larger networks. (c)  Synchronization order parameters ($R, R_U, R_V$) as a function of repulsive strength $K_R$.       
	$50$ independent numerical realizations are taken to obtain these observations of (a) and (b). Here $K_A=0.2$. ((d)-(f)) {\bf Impact of spanning tree in weak coupling.} The coupling parameters are set in oscillatory regime and  anti-phase is clearly established. (d) and (e) limit cycles and  time series of $x$ arbitrarily chosen nodes from the two disjoint subsets $U$ and $V$ in red and blue lines, respectively. (f)  Anti-phase synchronization between  
	$x_U$ and  $x_V$ for several arbitrarily chosen  nodes.
	Here, $K_A=0.01, K_R=-0.1$.}
	\label{fig2} 
\end{figure*} 

	
We aim here to find the partial negative control in a bipartite graph of limit cycle oscillators to obtain 
$F=0$
either in oscillatory regime or in steady state regime. The signature that we have got from our observed results (cf. Fig. \ref{fig1}) which directs us to  use the partial negative control in the bipartite graph with a proper arrangement, i.e., passing the negative links through a spanning tree embedded in the considered graph.
Therefore, we  hypothesize that  finding a spanning tree and passing the negative strength through it gives a faster zero frustration (anti-phase state)  compared  to randomly chosen negative links in a bipartite graph of limit cycle oscillators.

To validate our hypothesis, we start with two bipartite graphs having size 
$N=100$ with average degree $\langle k \rangle =12$ (Net1)
and  
$\langle k \rangle =6$ (Net2).  
We strategically 
select
$N-1$  
repulsive links in two different ways: 
(i) one set of links is chosen through a spanning tree  and (ii) other sets of links are chosen randomly.
The negative coupling strength 
($K_R$) 
of each branch in the tree
is changed continuously (adiabatically)
from 
$0$ to $-4$. 
We fix the coupling strength of each attractive link  at  
$K_A=0.2$.  
It is clear that the repulsive spanning tree  can induce zero frustration (Fig.\ \ref{fig2}(a)  with unfilled red circles for Net1 and unfilled black squares for Net2) in the bipartite graphs. However,  randomly chosen negative links cannot induce zero frustration (shown in Fig.\ \ref{fig2}(a)  with  filled  red circles (Net1) and filled black squares (Net2)). 
The result of getting 
$F=0$ 
in the bipartite network is consistent with previous work \cite{Levnajic2012} where an evolutionary process to construct non-frustrated ($F=0$) networks of phase repulsive dynamics for a certain upper limit on the number of edges is designed. On the other hand, in our study, only replacing a certain fraction of attractive links by negative ones can induce such phenomena.


Next, to study the effect of network size $N$, we   perform numerical simulations on eight networks ranging from $N=10$ to $N=2\times 10^3$ shown in Fig.\  \ref{fig2}(b). In all cases, calculated $F$ through random paths of length $(N-1)$ give  non-zero $F$, whereas if we pass the repulsive coupling through the branches of an underlying spanning tree, the numerically calculated $F$ is always close to zero. This confirms that our method is performing well even in large complex networks. $50$ independent statistical realizations are used to produce the Fig.\  \ref{fig2}(b). It should be noted that with increasing $N$, the basin of attraction becomes much constricted \cite{Levnajic2019} and hence, one should start with suitable initial conditions to enhance the probability of reaching $F=0$ through repulsive spanning tree. Another important observation of this figure is that the deviation of the calculated $F$ through randomly chosen edges of length $(N-1)$ is comparatively wide for larger $N$, as such dynamics can have multiple inhomogeneous steady states leading to a wide  range of  $F$ values, however they never reach to zero. 

It is clear now that nodes in same subset (either $U$ or $V$) will be in in-phase states, and they will be anti-phase to each other, which is further confirmed by the order parameters 
$Re^{i\psi}= \frac{1}{N} \sum_{j=1}^{N} e^{i\phi^j}$,
$R_Ue^{i\psi'}= \frac{1}{N_1} \sum_{l\in U} e^{i\phi^l}$ and 
$R_Ve^{i\psi''}= \frac{1}{N_2} \sum_{k\in V} e^{i\phi^k}$, 
where $\psi,\psi', \psi''$ are the common phases appear in the time of unison and $R, R_U, R_V$ are the degree of synchronization of  the whole set $\mathcal{V}=U~\cup~V$, $U$ and $V$, respectively. Here, the  number of nodes in the subsets $U$ and $V$ are  
$N_1$ and $N_2$, respectively and $N_1+N_2=N$. 
In Fig.\ \ref{fig2}(c) (for the network Net1), we observe that the order parameter in each subsets will follow in-phase solution 
($R_U\sim1, R_V\sim1$) 
at 
$K_R=-4.0$, where, as the global order parameter 
$R$ eventually reaches to zero, as phases of the two subsets follow anti-phase to each other.  This  feature 
($R\sim0$) 
is  consistent with 
$F\sim 0$. 
 Note that, the emergence of anti-phase between adjacent nodes 
 is calculated through the stationary phases obtained from steady states as it stabilizes the systems to stable fixed points. These fixed points are called as inhomogeneous steady states or oscillation death \cite{Hens_2013,OD1,OD2,OD3}. These coupling dependent fixed points manifest oscillation quenching by breaking the system’s symmetry. Interestingly, the spanning tree can do the same effect ($F$ and $R$ will be zero)  in limit cycle regime considered at lower negative strength 
 $K_R = -0.1$, when $K_A=0.01$. 
 We consider a sparse diluted bipartite graph having size  $N=100$ 
 with average degree 
 $\langle k\rangle=3$.
 Trajectories oscillate in limit cycle regime is depicted  with phase space diagram in Fig.\ \ref{fig2}(d).
  In Fig.\ \ref{fig2}(e), few time series are randomly chosen from the two subsets
  $U$ 
 (red color) and 
 $V$ (blue).
 The anti-phase synchronization manifold is confirmed from Fig.\ \ref{fig2}(f).
 
{ It is also clear that  a network having arbitrarily chosen degree sequences
	cannot always lead us into  anti-phase state
	($F\sim0$)
	 as the several graphs will not have bipartite structure.}
	The non-bipartite graph contains odd cycles and thus, the set of all vertices $\mathcal{V}$ can not be partitioned into two disjoint sets, i.e., the nodes within a subset can be connected to each other reflecting in-phase solution ($ f_{local}^{ij} > 0$). Therefore, $F$ will be deviated from the global minimum zero value. For instance, if one considers a  triangle, which is a non-bipartite graph and assigns all links as negative, the total $F$ can never be zero  \cite{Levnajic2012}.
	Let us assume the cardinality of the sets are respectively 
	$|U|=N_1$ and $|V|=N_2$. 
	Hence, the maximum number of possible links is $N_1N_2$, 
	so that the network remains bipartite. Using $N_1+N_2=N$, we get $g(N_1)=N_1(N-N_1)$, 
	which is maximized at 
	$N_1=\frac{N}{2}$ 
	and the maximum value of 
	$g$ is $\frac{N^2}{4}$. 
	So, any network can not be rearranged in a bipartite configuration, because at most $\frac{N^{2}}{4}$ links 
	are contained by a bipartite graph with
	$N$ nodes. 
	Therefore, the previous observation \cite{Levnajic2012} in the   restriction on the number of total edges in a network having zero frustration is due to the breaking of bipartite structure of a network.
 A natural question appears  {\it what will be the impact of negative spanning tree in a non-bipartite graph of limit cycle oscillators?}  Of course, the presence of odd-cycles in a non-bipartite graph will resist to set $F$ to zero. Although we will show (in sections VI and VII) that we can efficiently calculate the non-zero value of $F$ for any arbitrarily chosen non-bipartite graph and easily construct sparse non-bipartite graph for any non-zero desired frustration value ($F>0$).
We have avoided multistability (emergence of coexisting fixed points) \cite{Levnajic2011,Kaluza_Chaos_2010,Labavic2017} by adiabatically increasing the  negative coupling strength for all numerical calculations. 
A connected graph can have more than one spanning tree, but all spanning trees are connected  by
  $N-1$ edges and we would like to explore  the contribution of different repulsive  trees in  near future.
 To understand the stationary phase distributions of the steady states of LS limit cycle oscillators, we perform the bifurcation analysis of simple bipartite graph in the next section. For instance,   
Fig.\ \ref{fig1}(f) confirms that the entire network (a bipartite graph) reaches to steady states for strong negative coupling and follows some symmetric condition as the most of the fixed points are situated  either in second or fourth quadrant and the frustration index $F$ is zero. In the next section, we describe the bifurcation nature  for a small bipartite graph which will reflect why the nodes are in two opposite quadrants in the phase space. 
\\ 
 \section {Bifurcation analysis in a small  bipartite graph}
\begin{figure}[ht]
	\centerline{
		\includegraphics[scale=0.15]{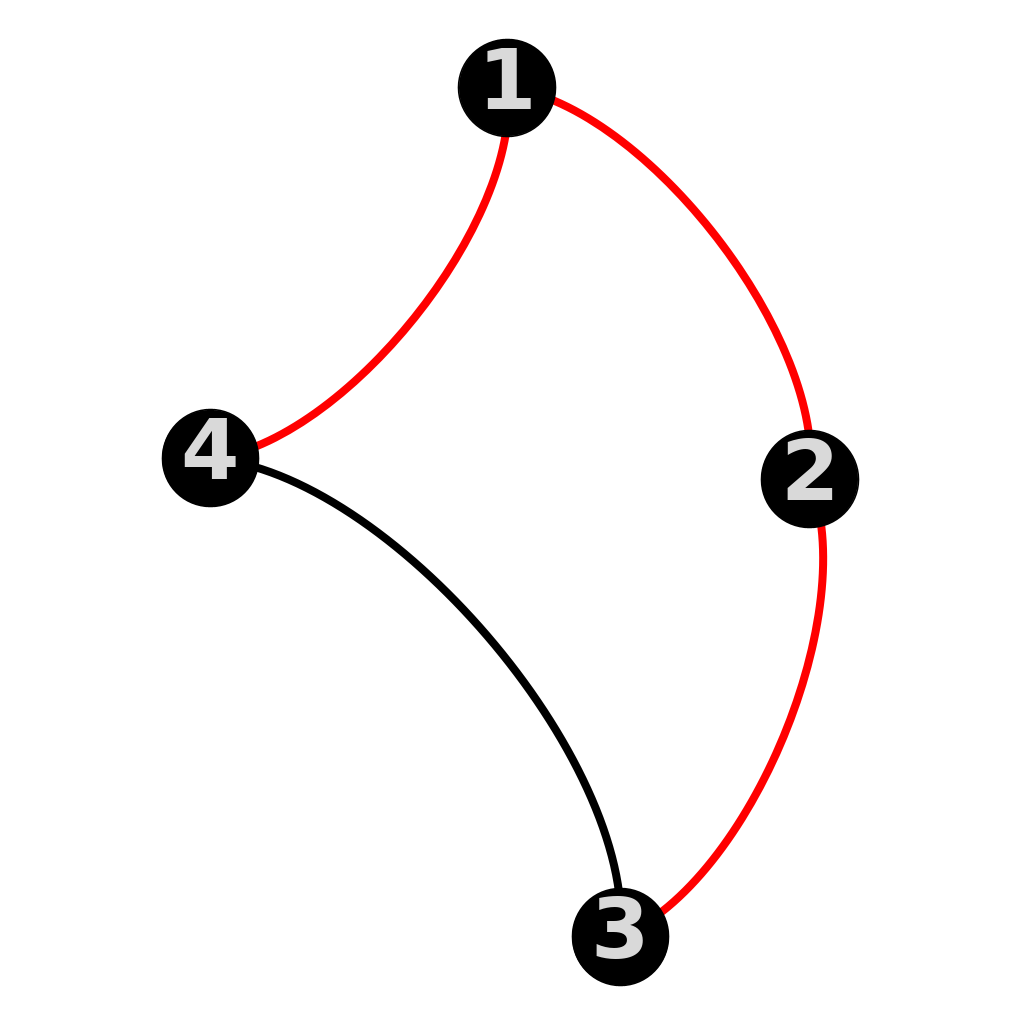} 	}
	\caption{{ {\bf Ring network of $4$ nodes.} Repulsive links are shown by red arcs and the black edge represents attractive interaction.
			Red links form  a spanning tree in the given graph (ring). Figure is drawn using the software Gephi \cite{Gephi}. }}
	\label{sfig1} 
\end{figure}

\begin{figure}[ht]
	\centerline{
		\includegraphics[scale=0.60]{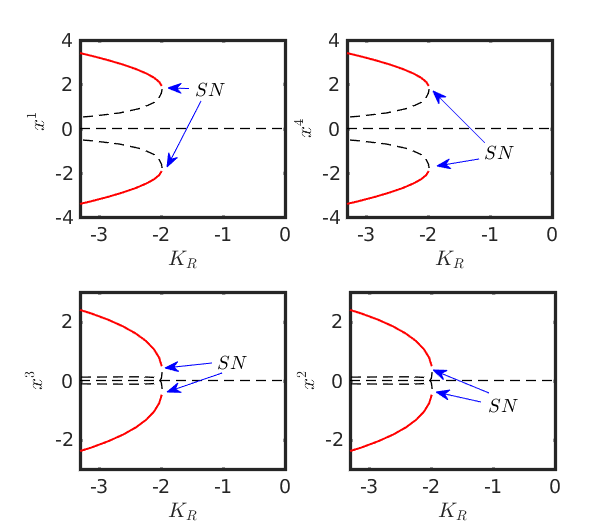} 	}
	\caption{{{\bf Bifurcation diagram of $N = 4$ coupled Landau-Stuart
				oscillators [using the network drawn in Fig. \ref{sfig1}] with $K_A=0.2$ and $\omega=3.0$.} Here, Red thick lines: stable fixed points; dashed black lines: unstable fixed points;  $SN$: saddle-node bifurcation.  }}
	\label{sfig2} 
\end{figure}

We consider a simple even cycle ring network  with 
$4$ nodes 
(Fig.\ \ref{sfig1}). A randomly chosen spanning tree inside the graph is shown by red arcs. 
We set the  attractive coupling strength at
$K_A = 0.2$ and  negative strength 
at $K_R=-4.0$.
These four coupled LS oscillators (Eqn.\ \ref{eq:1}) have 
$5$ 
real fixed points 
($FP_i:i=1,2,...,5$) in which
two of the fixed points (say $FP_1$: $x^1=-x^4=-3.827$, $y^1=-y^4=-0.8033$, $x^2=-x^3=2.797$,   $y^2=-y^3=1.057$, and $FP_2$: $x^1=-x^4=3.827$,  $y^1=-y^4=0.8033$, $x^2=-x^3=-2.797$,  $y^2=-y^3= -1.057$) 
 are stable nodes whereas the other three fixed points are unstable foci.
We are only interested to study the behavior of these two stable nodes. 
Noticeably, we observe that the derived stable fixed points $FP_1$ and $FP_2$ 
maintain a relation in the form
$x^1=-x^4$, $y^1=-y^4$ and $x^2=-x^3$, $y^2=-y^3$. 
Also for
$FP_1$, $(x^1,y^1)$ 
and 
$(x^3,y^3)$ 
are in the same quadrant (third quadrant in the phase space) whereas 
$(x^2,y^2)$ and 
$(x^4,y^4)$ 
stay in the first quadrant. This feature is  also applicable for 
$FP_2$ 
where 
$(x^1,y^1)$ and $(x^3,y^3)$ 
are in the first quadrant and the rest are in the third quadrant.  Clearly, the nodes which are disconnected, i.e., not neighbors to each other 
	 lie within same quadrant. This unique and interesting observation is consistent with the 
	 graph, as the nodes are split into two disjoint sets: 
$U=\{1,3\}$
and 
$V=\{2,4\}$. Note that, the system may undergo multistability, i.e., may change the value of $x^i$. However, the repulsive tree 
solely determine the sign of the stable fixed points. For instance (for a given initial condition) if  one gets
$x^1$ and $y^1$  
as positive (negative),
$x^3$ and $y^3$ 
will also be positive (negative).
The same features occur for 
$x^2$, $y^2$ and $x^4$, $y^4$.
This observation is also demonstrated with the help of the bifurcation diagram Fig.\ \ref{sfig2}.  Numerical bifurcation diagram of Fig.\ \ref{sfig2} using XPPAUT \cite{XPPAUT} yields that the stable nodes coexist, further if we increase 
$K_R$, 
the node comes closer to unstable fixed point and eventually annihilate around 
$K_R\sim-2.0$ through saddle-node bifurcation. In the next section, we calculate the stationary phases of all the nodes. Using the unique appearance of fixed points in the phase space, we establish a $0-\pi$ rule which further confirms the emergence of anti-phase state among neighboring nodes (i.e. zero frustration) for any bipartite graph.

\section {$0-\pi$ rule to determine the zero frustration in bipartite graph }
We  calculate the stationary phase value ($\tan \theta^i= \frac{y^i}{x^i}$) of each node in the given network (Fig.\ \ref{sfig1}) at 
$K_R=-4.0$. 
The stationary phase values up to $4$ decimal places are:
$\theta^1=-2.9347$, $\theta^2=0.3613$, $\theta^3=-2.7803$ and $\theta^4=0.2069$ of the nodes
$1$, $2$, $3$ and $4$, respectively. 
Though these phase values are not unique (can be altered for different initial conditions as noted in the bifurcation diagram Fig.\ \ref{sfig2}), but one significant property is observed:
the modulus of phase difference between two nodes (if link exists) is always 
$\pi$ 
for all the  links, therefore the frustration index
$F$
will be zero irrespective of that the initial conditions.
This will happen when
$|\theta^i-\theta^j|\simeq \pi ~~ \mbox{for} ~ \theta^i \in U=\{1,3\} ~ \mbox{and} ~ \theta^j \in V=\{2,4\}$, 
where 
$U$ 
and 
$V$
are disjoint sets 
($U \cap V =\phi$). 
We notice that  this inspection helps us to apply a change of origin in the $(r,\theta)$-plane 
without changing the directions of the axes. For instance, here the origin 
$(0,0)$ can be translated to $(0,-0.3612)$ 
so that in the new transformed co-ordinate system, we can acquire $\pi$ as the phase of each nodes of 
$U$ and phase $0$ 
is set for the vertices of $V$. 
\begin{figure}[ht]
	\centerline{
		\includegraphics[scale=0.20]{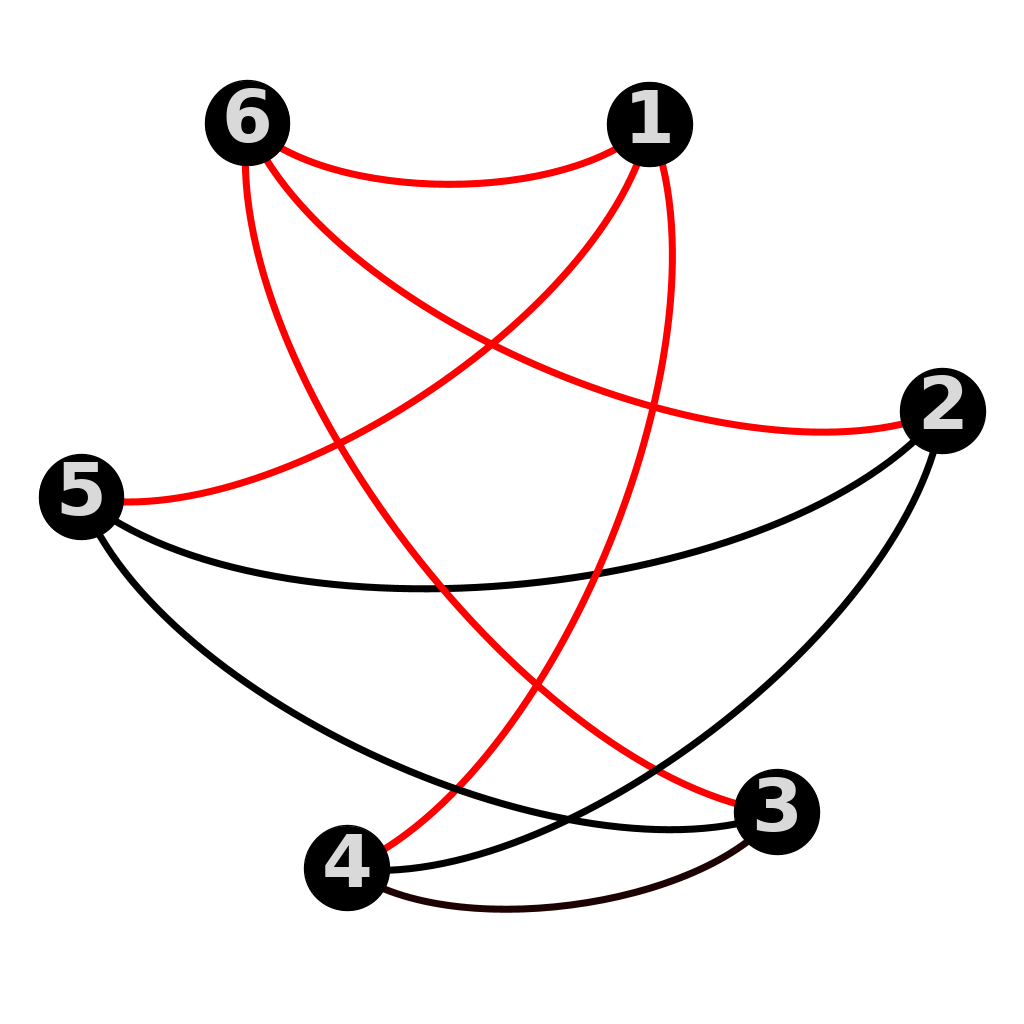} 	}
	\caption{{ {\bf Regular bipartite graph with six nodes.}  $K_{3,3}$ with $5$  repulsive links (red arcs) and 
	$4$ 
  attractive links (black arcs). Here 
  $U=\{1,2,3\}$ and 
  $V=\{4,5,6\}$. We assign 
  $0$ phase for $U$ set and 
  $\pi$  
for $V$ set (or vice versa). }}
	\label{sfig3} 
\end{figure}  
Similarly, for the utility graph 
$K_{3,3}$ 
(Fig.\ \ref{sfig3}), 
we find that the final phase values  for the chosen spanning repulsive tree are 
$\theta^1=0.1445$, $\theta^2=0.3401$, $\theta^3=0.3401$, $\theta^4=-2.8015$, $\theta^5=-2.8015$ and $\theta^6=-2.9971$, which again imply $|\theta^i-\theta^j|\simeq \pi ~~$ for $~ \theta^i \in U=\{1,2,3\} ~$ and $ ~ \theta^j \in V=\{4,5,6\}$ 
eventually determines 
$F$ 
as zero. 
Thus, with a suitable  phase shifting,  we can allocate $\pi$ 
as the phase of each nodes of 
$U$ and phase 
$0$ 
is set for the vertices of $V$ 
(or vice versa).
This whole observation (bifurcation analysis and obtained 
$\theta$ 
values for each node) enables us to establish a 
$0-\pi$ 
rule  for any arbitrary  bipartite graph
to predict
 $F$ 
without investigating the  local dynamical units at higher negative strength.
Note that, this  rule is also verified for Figs.\ \ref{fig1}(f),
 \ref{fig2}(b) and \ref{fig2}(d) 
 (for sufficient repulsive strength in three bipartite graphs)  
 in which any arbitrary node will stay in anti-phase with respect to its nearest neighbors.  
 In this context, the local frustration
 $f^{ij}_{local}$
 will be zero as $|\theta^j-\theta^i|=\pi$ 
 which further implies  
 $F=0$.

\section{ Designing a non-bipartite  network for desired frustration}

The unique feature of  
$F$  
having zero value   for bipartite graph elucidates us to formulate a design principle of constructing a non-bipartite graph having desired (given) frustration 
$F_{desired}$ 
for a given tree.  As discussed above,   the disjoint nodes of a spanning tree  will be distributed with two-phase values: either 
$0$  (say for $U$) or 
$\pi$ (say for $V$). 
Now, the additional attractive link randomly attached to any two nodes of the tree may give us two values of  local frustration: zero, if both of them have different phases to each other (two nodes are not chosen from the same set, i.e., one is chosen from 
$U$ and the other one is from 
$V$) or
 $2$, 
if two nodes either in zero phase or in 
$\pi$ phase (nodes are chosen from the same set, i.e., either both are from $U$ or from $V$).
 Assume, we have a spanning tree of  
 $N$ 
 nodes with negative links  and we would like to construct a network such that we will eventually reach a desired frustration 
 ($F_{desired}$) 
index. For this, suppose we have to add  
 $m$ (say)
number of minimum links such that we can reach 
$F_{desired}$. Therefore, the underline condition is 
  $F_{desired}\sim\frac{2m}{L}$.
Here $L$ 
 is the total number of links in the network, i.e.,
 $m+(N-1) = m+N_L= L$.
 The number 
 $2$ appears in numerator due to the contribution of positive links in the local frustration term in the time of network construction.
In Appendix, we describe each step in details  for the  construction of our  network. Note that, the algorithm is based on the    following  hypothesis that
 {\it a spanning tree will have two phase values : 
 $0$ for $U$  and $\pi$ for $V$ 
 or vice versa and  link will be added between two nodes if and only if their phase difference is zero}.
 Based on this hypothesis, we have added links (between the nodes in the tree) continuously unless we reach our desired $F$.
  The required number of attractive links 
  ($m$) 
  for a given 
  $F$ 
  are shown in the Fig.\ \ref{fig3}. 
 The 
 $F_{desired}$ of the newly constructed network 
 perfectly matches with the numerically calculated 
 $F$ 
   by considering LS dynamics over the top of the 
   same network (shown with yellow circles). 
   In conclusion, we can state that for a given tree one can  easily construct a network by adding few edges  which can give us our desired frustration index $F$. 
   The approach is simple and it creates a little dense network from the highly diluted graph (a tree only) to shift the 
   $F$ 
   value from zero to a desired one. {Finally, we obtain the relation between $F_{desired}$ and $m$ as follows
    \begin{equation}
m= \ceil[\Bigg]{\frac{(N-1)F_{desired}}{2-F_{desired}}},~ 0.0 \leq F_{desired} < 2.0.
\label{equ4}
\end{equation} 
  By suitable isometric translation operator in the Euclidean space, one can reach 
   $XY=C^2$, where $X=2-F_{desired}$, $Y=m+(N-1)$ and $C^2=2(N-1)$. It is clear that $m$ and $F_{desired}$  follow a  hyperbolic growth (for a given $N$) to each other, which is clear from the Fig.\ \ref{fig3}.
Note that, we need to add more number of attractive links to obtain higher $F_{desired}$
and 
our algorithm always predicts the lower most $m$ for a desired $F$. 
     One can obtain exact in-phase synchronization, i.e.\, ($F=2$), 
   the network needs significantly large number of positive links 
   ($m$) 
   to completely diminish the effect of negative strength connected through the tree.  
\begin{figure}[ht]
	\centerline{
    \includegraphics[scale=0.43]{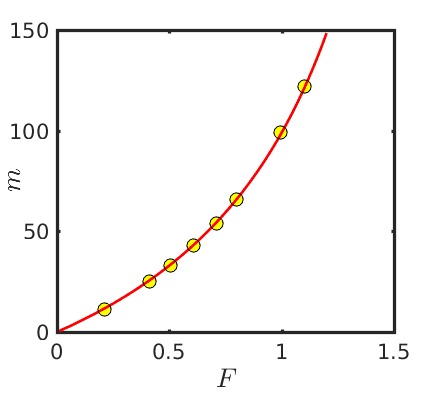} 	}
	\caption{{\bf Network construction from a given spanning tree with $\bf F_{desired}$.} Red line represents the analytical curve as per our proposed mechanism  (based on $0-\pi$ proposition) for network construction with the help of given $F$ and a given spanning tree of length $N-1=99$. Yellow circles are the verification by assigning Landau-Stuart oscillators as the local unit dynamics on the top of our constructed network. }
	\label{fig3} 
\end{figure}

   However, our algorithm creates a strong restriction on the upper bound of   
   $F$ in  a network of 
   $N$
   oscillators.  If $N$ number of nodes connected in a negative spanning tree, then we can add at most $2$ ${{N/2} \choose 2}$ number of attractive links. The reason is that a bipartite graph will have maximum $\frac{N^2}{4}$ links if the disjoint sets are equally divided which is discussed 
    earlier. This information further  restricts on  the upper bound of $F_{desired}$. For instance, $m=2{{N/2} \choose 2} = 2450$ number of attractive links can be added as per our algorithm for any network of $N=100$ oscillators. Hence, maximum $F_{desired}= \frac{2m}{m+(N-1)}~=~\frac{4{{N/2} \choose 2}}{2{{N/2} \choose 2}+(N-1)} ~ \simeq 1.92$ can be obtained.
~In fact, this is the largest upper bound of $F_{desired}$ 
with the presence of repulsive spanning tree, as one can add more links ($>2{{N/2} \choose 2}$) by connecting two disjoint sets $U$ and $V$,
which are in anti-phase among themselves. Therefore, these additional edges do not contribute in $f_{local}^{ij}~(=0)$,
 but those links increase $L$ and consequently, reduce 
 $F$. As a conclusion, we can never reach to complete in-phase synchronization, i.e., $F_{desired}=2$ as our analytical expression given in Eq.\ (\ref{equ4}) 
 has singularity at $F_{desired}=2$.
   }
 \section{ Calculation of $F$ for arbitrary network} The constructed non-bipartite graph described in the last section has two types of links: repulsive links in the spanning tree contribute zero in the frustration index 
$F$, and the $m$  attractive links (edges within the same set $U$ or $V$)
only contribute on 
$f_{local}^{ij}$.
However, an arbitrary chosen graph (e.g. networks used in the Fig.\ \ref{fig2}(a)) will also have links which may not contribute in the local frustration index 
$f^{ij}_{local}$ 
as those additional links connect random nodes between the subsets 
$U$ and $V$. 
The maximum  number of these types of links  will be $\frac{N^2}{4}-(N-1)$, 
i.e., the difference between the maximum number of  bipartite links possible in the network and a spanning tree which is a part of that bipartite graph. 
However, these inactive links may create multistability \cite{Kaluza_Chaos_2010} in 
$0-\pi$ 
states of the nodes and may perturb the value of $f^{ij}_{local}$  and hence changes the value of $F$. 
To test our $0-\pi$ proposition in the considerably dense networks,
 we have designed  several small graphs of size 
$6$. We also check how different spanning trees as repulsive links work inside a graph.

\begin{figure*}[ht]
	\centerline{
    \includegraphics[scale=0.25]{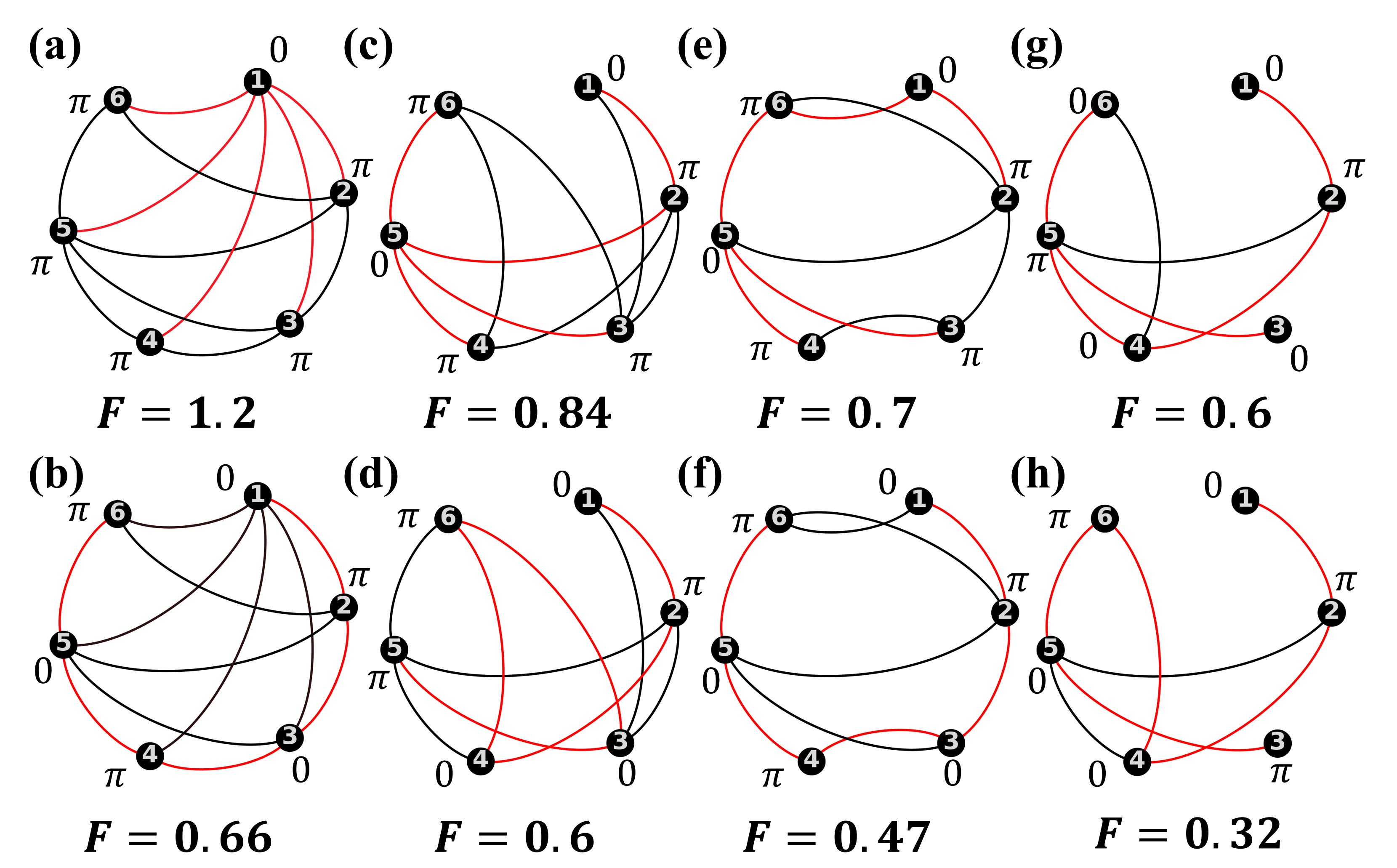} 	}
	\caption{{{\bf Numerically calculated $F$  for $8$ different networks with size {\bf $N=6$}.} Each column represents the same network architecture containing different spanning trees. Branches of the tree are represented using red arcs and the attractive links are shown using black arcs. Using bipartiteness of the tree, $0$ and $\pi$ phases are assigned to each node of the network. Figures are drawn using the software Gephi \cite{Gephi}. ($K_A=0.2$ and $K_R=-4.0$)  }}
	\label{sfig4} 
\end{figure*}

We  start  with four different networks of $6$ nodes with total number of links:
(i)  $12$ links  (first column),  
(ii) $10$ links  (second column), 
(iii) $9$ links (third column) and
(iv) $7$ links (fourth column) in Fig.\  \ref{sfig4}. 
We have identified two separate spanning  trees for each network with different links. The branches of  spanning trees for each network are shown with red arcs and the attractive links  in black arcs. Each of the spanning trees will have 
$N-1=5$ 
links irrespective of the variation of link density. However, the same network may reveal  different 
$F$  
which solely depends on the choice of spanning tree and it's attractive counterpart. 

\begin{figure}[ht]
	\centerline{
		\includegraphics[scale=0.5]{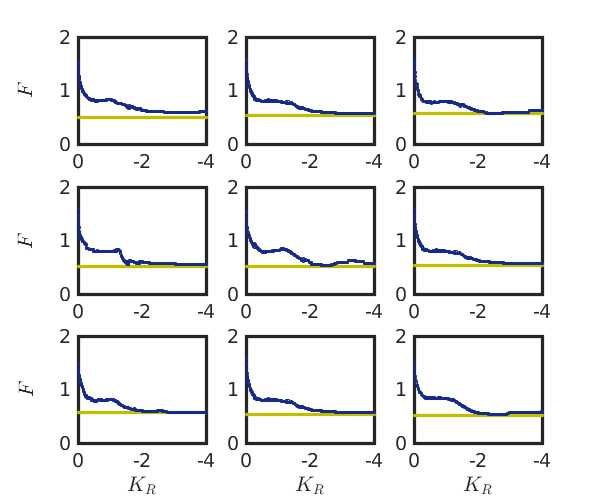} 	}
	\caption{{\bf Theoretical prediction of $F$ based on $\bf 0-\pi$ proposition.} Prediction of $F$ for a network of $N=100$ nodes with average degree $\langle K \rangle =4$. Here $9$ different spanning trees are considered. Yellow line is the analytically predicted $F$ and blue curve is the numerically obtained $F$.  }
	\label{fig4} 
\end{figure}

For instance,   (i) Figs.\ \ref{sfig4}(a) and \ref{sfig4}(b) have 12 links in which the   Fig.\ \ref{sfig4}(b)   gives lesser frustrated index 
($F=0.66$) 
than the other choice of a spanning tree shown in Fig.\ \ref{sfig4}(a) in which 
$F=1.2$. 
The same type of features is revealed for   
$L=10$ 
shown in
Figs.\ \ref{sfig4}(c) and \ref{sfig4}(d). Two trees are generated and a comparison reflects that 
$F$ 
becomes lesser for a spanning which possess larger diameter. We like to explore the impact of different trees in details for large network in near future.
We have also used more diluted graphs  where 
$L=9$ (Figs.\ \ref{sfig4}(e) and \ref{sfig4}(f)) and $L=7$ (Figs.\ \ref{sfig4}(g) and \ref{sfig4}(h)). Next, we calculate the  frustration value of the entire network utilizing the $0-\pi$ rule mentioned above.  
 As we have shown that a network can reveal zero frustration {\it if and only if} the network is bipartite
and  a spanning-tree inside the graph is required to pass the negative signals adiabatically. We assume here, that the underlined tree (which itself is bipartite)  will always lead  to zero frustration. For instance,
the small graph considered in Fig. 
\ref{sfig4}(a), we assign the local $\theta$ values (through the considered tree shown by red lines) on the vertices as follows:
$V^1=0,V^2=\pi,V^3=\pi,V^4=\pi,V^5=\pi$ and $V^6=\pi$. 
Therefore, the  frustration of the tree will be  zero which can be calculated by accumulating all the local (link) frustration values
$f^{1,6}+f^{1,2}+f^{1,3}+f^{1,4}+f^{1,5}=0$. 
Based on this assumption,
the total frustration in the entire network appears from the sole contribution of attractive links is
$F=\frac{f^{2,6}+f^{5,2}+f^{5,3}+f^{5,4}+f^{3,4}+f^{2,3}+f^{5,6}}{12}=\frac{2+2+2+2+2+2+2}{12} \simeq 1.1666$. 
In a similar manner, we calculate here the total frustration (from the attractive links) of the graph represented in  Fig.\ \ref{sfig4}(b) by setting the 
$\theta$ values of the vertices of the spanning trees
$V^1=0,V^2=\pi,V^3=0,V^4=\pi,V^5=0,V^6=\pi$. 
Therefore, the total frustration  of entire network will be like
$F=\frac{f^{1,6}+f^{1,3}+f^{1,4}+f^{1,5}+f^{2,5}+f^{2,6}+f^{3,5}}{12}=\frac{0+2+0+2+0+2+2}{12} \simeq 0.67$.
The numerically calculated (by considering LS dynamics over the top of the network)
$F$
 values  mentioned in Figs.\ \ref{sfig4}((a)-(h)) are all closely matched with our 
 calculation based on 
 $0-\pi$ 
 rule.

 However as we have mentioned earlier, these types of dynamical graphs are multistable by nature which may give different $F$ index of graph
 $G$ \cite{Levnajic2011,Levnajic2012}.
To avoid such spurious effect, we have calculated the 
 $F$ 
index by adiabatically increasing the negative strength
 unless the total frustration value saturates in a specific value. 
 Note that,  most of these graphs contain the inactive links (as corresponding $f^{ij}_{local}=0$),  but the numerical results fit with our 
$0-\pi$ proposition for appropriate high negative coupling strength.  
We have also checked this phenomenon in a network of $100$ 
 oscillators (shown in Fig.\ \ref{fig4}).
  Diverse spanning trees (we have traced $9$
 unique trees) have been chosen for that static network.  
 The yellow lines are the theoretically predicted $F$ and the blue lines obtained numerically. In higher negative strength ($K_R\simeq -4$), the theoretical value of $F$ closely match with the numerically calculated $F$. 

\section{Conclusions}
Our work undermines that a group of amplitude oscillators placed over a bipartite network can lead us to anti-phase states ($F=0$) if the negative  links are passed through the branches of a  spanning tree. Noticeably,  we  establish that  such partial use of repulsive interactions on undirected complex networks helps to achieve the highest phase difference of 
$\pi$ 
between two adjacent amplitude oscillators depending on network topology.
  This negative control in a bipartite graph is simple, elegant and  cost effective which can outperform the other random choices of negative  links.
Next, we investigate the role of spanning trees in  a bipartite graph of limit cycle  in a broader level:  we construct a $0-\pi$ rule to understand the zero frustration. We also uncover the  
construction of a weakly dense non-bipartite graph for the desired $F$ with the help of $0-\pi$ rule. 
Finally, we have shown that the negative spanning tree of any non-bipartite  graph can successfully determine the non-zero $F $ value. 
 Our approach  can be easily implemented to other limit cycle and Kuramoto dynamics. Also, we can implement the attractive-repulsive technique in phase-frustrated Kuramoto oscillator \cite{Nicosia2013} in which the mixed coupling  can be constructed through the heterogeneous phase-frustrated values contained within the coupling function. We would like to explore the impact of phase lag to generate anti-phase states in the near future. Our investigation also points out to the fact that zero frustration $(F = 0)$, {\it i.e.}\ anti-phase synchronization emerges only in the bipartite network, whereas the non-bipartite network reveals positive $F$ value irrespective of the setting of negative links within the considered graph. Thus, the dynamical measure $F$ serves as a unique fingerprint to distinguish between bipartiteness and non-bipartite structure of a network. In our designed bipartite network of attractive-repulsively coupled amplitude-oscillators, oscillators tend to organize their phase values as colors in a solution of $2$-colorable graph, which is validated using several bipartite networks of various sizes and different average degrees throughout the paper. The study may unfold the coherent and de-synchronization phenomena broadly in an environment of attractive-repulsive interaction.  Our perceived study may be helpful to understand underlying frustrated dynamical behavior of neuronal networks \cite{Li2011}.


\section{ACKNOWLEDGEMENTS}
The authors gratefully acknowledge the anonymous referees for their insightful suggestions that helped in considerably improving the manuscript. SNC and DG are supported by the Department of Science and Technology, Government of India (Project No. EMR/2016/001039). CH is supported by DST-INSPIRE Faculty Grant. No. IFA17-PH193.

\section{APPENDIX: Algorithm for designing suitable network for desired frustration based on $0-\pi$ rule}
Suppose, the value of $F$ at the steady state is given for a network and the configuration of the underlying repulsive tree is also supplied to us, though the exact architecture of the original network of $N$ nodes is unknown. So, with the help of that two specified information, which are namely (i) the spanning tree of that network and (ii) the value of $F$, we would like to design network. Then one can address the question in the following way, \textit {how to construct a network of $N$ nodes having the congruent underlying tree, which can give us the desired $F$? To proceed further, we initiate an algorithm in such a fashion that it will help us to create a new network of 
$N$ 
nodes containing the same underlying spanning tree which will adapt itself by adding appropriate  links until the  desired frustration 
$F$
 is reached.}\\ 
 
The steps of algorithm are depicted below precisely:-

\begin{enumerate}

\item $F_{desired}$ and a spanning tree of a network with $N$ nodes are given as inputs.

\item The vertex set of the spanning tree is decomposed into  two disjoint and independent sets $U$ and $V$. \vspace{0.2cm}

\item We assign $\pi$ as the phase of each node of $U$ and phase $0$ for the vertices of $V$ (or vice versa). 

\item Two nodes $i$ and $j$ are chosen arbitrarily either from the set $U$ or from $V$, so that there is no link in between those nodes at prior.

\item A link is added between these two chosen nodes $i$ and $j$, and $F$ is calculated for the constructed network using our $0-\pi$ proposition. 


\item If $F<F_{desired}$, then go to step $(4)$ and repeat the process.
Otherwise,  the process is terminated.
\end{enumerate}

 Our constructed network is capable of giving rise to the given $F_{desired}$, which we verify by assigning Landau-Stuart oscillators as the local unit dynamics on the newly constructed network and integrate it with Runge-Kutta-Fehlberg method with an error estimator of order O($h^5$), where $h=0.01$ is the step length.  
We avoid the multistability effect by adiabatically changing the negative strength from $0$ to $-4$
until the  
$F$
is reached to asymptotically saturated value.
We have collected the numerical frustration index $F$ from the saturated domain to match with our theoretical prediction. 

To attain the given $F_{desired}$ with the given spanning tree using our algorithm, we need total number of attractive links  $m$, which satisfies the following relation,

$$F_{desired}=\frac{2m}{L}~,$$ i.e.,
$$ F_{desired}=\frac{2m}{m+(N-1)}~, $$which gives
\begin{equation}
m=\ceil[\bigg]{\frac{(N-1)F_{desired}}{2-F_{desired}}}, 0.0 \leq F_{desired} < 2.0~.
\end{equation}

\end{document}